\documentclass[aps,prd,showpacs,showkeys,hyperref]{revtex4}
\usepackage{graphicx}

\usepackage{amsmath}
\usepackage{amssymb}

\usepackage{bbm}

\DeclareFontFamily{OT1}{rsfs}{}
\DeclareFontShape{OT1}{rsfs}{m}{n}{ <-7> rsfs5 <7-10> rsfs7 <10->rsfs10}{}
\DeclareMathAlphabet{\mycal}{OT1}{rsfs}{m}{n}
\newcommand{\lan}{{\mycal L}}

\newcommand{\p}{\partial}

\newcommand{\vp}{\mathbf{p}}

\newcommand{\dual}[1]{\overset{{}^{{}^{\boldsymbol{\neg}}}}{\smash[t]{#1}}}
\newcommand{\gdual}[1]{\overset{\:{}^{{}^{\boldsymbol{\neg}}}}{\smash[t]{#1}}}

\begin{document}

\title{Dark spinors with torsion in cosmology}

\author{Christian G.~B\"ohmer}
\email{c.boehmer@ucl.ac.uk}
\affiliation{Department of Mathematics, University College London,
             Gower Street, London, WC1E 6BT, UK}

\author{James Burnett}
\email{j.burnett@ucl.ac.uk}
\affiliation{Department of Mathematics, University College London,
             Gower Street, London, WC1E 6BT, UK}

\date{Draft of \today}

\begin{abstract}
We solve one of the open problems in Einstein-Cartan theory,
namely we find a natural matter source whose spin angular momentum
tensor is compatible with the cosmological principle. We analyze the
resulting evolution equations and find that an epoch of accelerated
expansion is an attractor. The torsion field quickly decays in that
period. Our results are interpreted in the context of the standard
model of cosmology.
\end{abstract}

\pacs{04.50.+h, 04.40.-b}
\keywords{Einstein-Cartan theory, dark spinors, torsion}

\maketitle

\section{Introduction}

General relativity is a successful theory in agreement with a vast number of observations. It is based on the Einstein-Hilbert action which yields the field equations if varied with respect to the metric. If, however, the metric and the connection (more precisely the non-Riemannian part of the connection) are considered as {\it a priori} independent variables, two field equations are obtained. The first one relates the Einstein tensor (not necessarily symmetric) to the canonical energy-momentum tensor, while the other field equation relates the skew-symmetric part of the connection, the torsion tensor, to the spin angular momentum of matter, see e.g.~\cite{Hehl:1973,Hehl:1974,Hehl:1976kj,Hehl:1995ue,Hammond:2002rm,Trautman:2006fp}. Spin and torsion are related by algebraic equations, and torsion vanishes in the absence of sources.

The cosmological principle states that the universe is homogeneous and isotropic on very large scales. More mathematically speaking  the four dimensional spacetime $(M,g)$ is defined by $3d$ space-like hypersurfaces of constant time which are orbits of a Lie group G action on $M$, with isometry group $SO(3)$. We assume all fields to be invariant under the action of G which means $\mathcal{L}_{\xi}g_{\mu \nu} = 0$ and $\mathcal{L}_{\xi}T_{\mu \nu}{}^{\lambda} = 0$ where $\mathcal{L}_{\xi}$ denotes the Lie derivative with respect to the generator of the group. This assumption reduces the cosmological metric to the well know Friedman-Lema\^{\i}tre-Robertson-Walker form which is characterized by the scale factor and the geometry of the constant time hypersurfaces. If applied to the torsion of spacetime, it reduces the components compatible with the cosmological principle to a spatial axial torsion and a vector torsion part~\cite{Tsamparlis:1979}.

Cosmological models with torsion were pioneered by Kopczy\'nski in~\cite{Kopczynski:1972,Kopczynski:1973}, who assumed a Weyssenhoff fluid~\cite{Weyssenhoff:1947} to be the source of both curvature and torsion. The cosmological principle was first extended to Einstein-Cartan theory in~\cite{Tsamparlis:1979}, where it was also suggested to reconsider the results in~\cite{Kopczynski:1972,Kopczynski:1973}, since the Weyssenhoff fluid turns out to be incompatible with the cosmological principle (see also~\cite{Obukhov:1987yu,Boehmer:2006gd,Brechet:2008tz}). An elaborate analysis of the most general action up to quadratic terms in curvature and torsion assuming the cosmological principle can be found in~\cite{Goenner:1984}. Analytical solutions of the Riemann-squared gravity have recently been discussed in a cosmological context in~\cite{Lasenby:2005nw}. Non-Riemannian models of cosmology in general have been discussed in~\cite{Puetzfeld:2004df,Puetzfeld:2004yg,Puetzfeld:2004sw,Puetzfeld:2007ye}

However, nobody has so far succeeded in constructing a non-trivial spin angular momentum tensor in cosmology by minimally coupling matter fields to the geometry. We show that the minimally coupled eigenspinors of the charge conjugation operator~\cite{jcap,prd} yield a spin tensor compatible with the cosmological principle.

These spinors belong to a wider class of so-called flagpole spinors~\cite{daRocha:2005ti}. They are non-standard spinors according to the Wigner classification and obey the unusual property $(CPT)^2=-\mathbbm{1}$. Hence, their dominant coupling to other fields is via the Higgs mechanism or via gravity~\cite{jcap,prd}. The particles associated with such a field theory are naturally dark and we will refer to them as dark spinors henceforth, note that they were originally named Elko spinors.

Dark spinors are defined by
\begin{align}
      \lambda = \begin{pmatrix} \pm \sigma_2 \phi^{\ast}_{L} \\
                \phi_L \end{pmatrix} \,,
      \label{eq:i1}
\end{align}
where $\phi^{\ast}_{L}$ denotes the complex conjugate of $\phi_L$ and $\sigma_2$ denotes the second Pauli matrix. For a detailed treatment of the field theory of the eigenspinors of the charge conjugation operator we refer the reader to~\cite{jcap,prd}. Dark spinors have an imaginary bi-orthogonal norm with respect to the standard Dirac dual $\bar{\psi} = \psi^{\dagger}\gamma^0$, and in order for a consistent field theory to emerge the dual is given by
\begin{align}
       \dual{\lambda}_u = i\,\varepsilon_u^v \lambda_v^{\dagger} \gamma^0 \,,
       \label{eq:i3}
\end{align}
with $\varepsilon_{\{+,-\}}^{\{-,+\}}=-1=-\varepsilon_{\{-,+\}}^{\{+,-\}}$ such that
\begin{align}
      \dual{\lambda}_u(\vp) \lambda_v(\vp) = \pm\, 2m\, \delta_{uv} \,,
      \label{eq:i4}
\end{align}
where $\vp$ denotes the momentum.

Due to their formal structure, dark spinors couple differently to gravitation than scalar fields or Dirac spinors~\cite{Boehmer:2006qq}, eigenspinors of the parity operator. This allows for many interesting applications, for instance, in~\cite{Boehmer:2007ut} it has been shown that dark spinors naturally yield an anisotropic expansion in the context of cosmological Bianchi type I models. This allows for a suppression of the low multipole amplitude of the primordial power spectrum. The primordial power spectrum of the quantum fluctuations of dark spinors has been investigated in~\cite{Boehmer:2008rz,Gredat:2008qf} where is was found that the small scale power spectrum essentially agrees with that of scalar field inflation while the large scale power spectrum shows new features.

The paper is organized as follows. In Section~\ref{sec:ec} we briefly summarize Einstein-Cartan theory. In Section~\ref{sec:cos} the cosmological field equations in Einstein-Cartan theory in the presence of dark spinors are presented. We analyze the field equations qualitatively and numerically in Section~\ref{sec:dyn} and conclude our work in Section~\ref{sec:con}.

\section{Einstein-Cartan theory with dark spinors}
\label{sec:ec}

It is well known that in general relativity the presence of matter yields the metric energy momentum tensor that acts as the source term of the curvature of spacetime. From a field theory point of view, this tensor is related to translations. However, special relativity is based on the Poincar\'{e} Lie algebra which has two conserved quantities, mass squared and angular momentum squared. It is therefore natural to extend the theory of general relativity by taking into account the spin angular momentum as an additional property of matter, which is associated to rotations.

In line with standard conventions we use Greek letters $(\mu,\nu,\ldots)$ taking values $(t,x,\ldots)$ for holonomic indices and Latin letters $(a,b,\ldots)$ taking values $(\hat{0},\hat{1},\ldots)$ for anholonomic indices. The line element is defined by
\begin{align}
      ds^2 = g_{\mu\nu}\, dx^{\mu} dx^{\nu} = g_{ab}\, e_{\mu}^a e_{\nu}^b\, dx^{\mu} dx^{\nu}.
      \label{eq1-anhol to hol}
\end{align}
We assume there exists a metric compatible covariant derivative operator $\tilde{\nabla}_{\lambda} g_{\mu\nu} = 0$, and do not require the connection to be symmetric. The torsion tensor is defined as the skew-symmetric part of the (full) connection
\begin{align}
      T_{\mu\nu}{}^{\lambda} = \tilde{\Gamma}^\lambda_{[\mu\nu]} =
      \frac{1}{2} \bigl(\tilde{\Gamma}^\lambda_{\mu\nu}-
      \tilde{\Gamma}^\lambda_{\nu\mu} \bigr),
      \label{eq:t1b}
\end{align}
and in turn we can decompose the connection into the Christoffel symbol and an additional piece, the contortion tensor
\begin{align}
      \tilde{\Gamma}^\lambda_{\mu\nu} = \Gamma^\lambda_{\mu\nu} - K_{\mu\nu}{}^{\lambda}.
      \label{eq:t1a}
\end{align}
Combining the latter relation with the definition of torsion~(\ref{eq:t1b}), it follows that torsion and contortion are algebraically related by
\begin{align}
      T_{\mu\nu}{}^{\lambda} = \frac{1}{2} (K_{\nu\mu}{}^{\lambda}-K_{\mu\nu}{}^{\lambda}).
\end{align}

The covariant derivative when acting on a spinor is defined by
\begin{align}
      \tilde{\nabla}_a \lambda = \partial_a \lambda - \frac{1}{4} \Gamma_{a} \lambda +
      \frac{1}{4} K_{abc} \gamma^b \gamma^c \lambda,
      \label{eq:t1}
\end{align}
where $\Gamma_a=\Gamma_{abc}f^{bc}$ and $f^{bc} = \frac{1}{4} [\gamma^b,\gamma^c]$ are the generators of the Lorentz group. The anholonomic and holonomic connections are related by
\begin{align}
      \Gamma^{c}_{ab} = e^{\mu}_{a} e^{\nu}_{b} \nabla_{\mu} e_{\nu}^{c} =
      e^{\mu}_{a} e^{\nu}_{b} e^{c}_{\lambda} \Gamma_{\mu\nu}^{\lambda}
      - e^{\mu}_{a} e^{\nu}_{b} \p_{\mu }e^{c}_{\nu},
      \label{eq:d4b} \\
      \Gamma_{abc} = g_{cd} \Gamma_{ab}^{d}, \qquad \Gamma_{a(bc)}=0.
      \label{eq:d4c}
\end{align}
We use the $\gamma$-matrices in the chiral form
\begin{align}
      \gamma^{\hat{0}} = \begin{pmatrix} \mathbb{O} & \mathbbm{1} \\
        \mathbbm{1} & \mathbb{O} \end{pmatrix}, \qquad
      \gamma^{n} = \begin{pmatrix} \mathbb{O} & -\sigma^i \\
        \sigma^i & \mathbb{O} \end{pmatrix}, \qquad
      \gamma^{\hat{5}} = \begin{pmatrix} \mathbbm{1} & \mathbb{O} \\
        \mathbb{O} & -\mathbbm{1} \end{pmatrix}, \qquad
      n=\hat{1},\hat{2},\hat{3}.
      \label{eq:d2}
\end{align}
with $\gamma^{\hat{5}} = i\gamma^{\hat{0}}\gamma^{\hat{1}}\gamma^{\hat{2}}\gamma^{\hat{3}}$, and the $\gamma$-matrices satisfy
\begin{align}
      \gamma^{(a} \gamma^{b)} = g^{ab}, \qquad
      \gamma^{\mu} = e^{\mu}_{a} \gamma^a, \qquad
      \gamma^{(\mu} \gamma^{\nu)} = g^{\mu\nu}.
      \label{eq:d2a}
\end{align}

The action of Einstein-Cartan gravity is
\begin{align}
      S = \int \Bigl( \frac{M_{\rm pl}^2}{2} \tilde{R} + \tilde{\lan}_{\rm mat} \Bigr)
      \sqrt{-g}\, d^4 x,
      \label{eq:a1}
\end{align}
where $\tilde{R}$ is the Ricci scalar computed from the complete connection with contortion contributions, $g$ is the determinant of the metric, $\tilde{\lan}_{mat}$ denotes the matter Lagrangian and $1/M_{\rm pl}^2 = 8\pi G$ is the coupling constant; the speed of light is set to one $(c=1)$. The resulting field equations are
\begin{align}
      \tilde{G}_{ij} = \tilde{R}_{ij} - \frac{1}{2} \tilde{R} g_{ij} &= \frac{1}{M_{\rm pl}^2}\,\Sigma_{ij},
      \label{eq:t8} \\
      T^{ij}{}_{k} + \delta^i_k T^{j}{}_{l}{}^{l} - \delta^j_k T^{i}{}_{l}{}^{l}
      &= M_{\rm pl}^2\,\tau^{ij}{}_{k},
      \label{eq:t9}
\end{align}
where $\tau^{ij}{}_{k}$ is the spin angular momentum tensor, defined by
\begin{align}
      \tau_{k}{}^{ji} = \frac{\delta \tilde{\lan}_{\rm mat}}{\delta K_{ij}{}^{k}},
\end{align}
and $\Sigma_{ij}$ is the total energy-momentum tensor
\begin{align}
      \Sigma_{ij} = \sigma_{ij} + (\tilde{\nabla}_k-K_{lk}{}^{l})
      (\tau_{ij}{}^{k}-\tau_{j}{}^{k}{}_{i}+\tau^{k}{}_{ij}),
      \label{eq:t10}
\end{align}
where $\sigma_{ij}$ is metric energy-momentum tensor
\begin{align}
      \sigma_{ij} = \frac{2}{\sqrt{-g}} \frac{\delta(\sqrt{-g}\tilde{\lan}_{\rm mat})}{\delta g^{ij}}. 
\end{align}
The field equations~(\ref{eq:t9}) are in general 24 algebraic equations, and in the absence of spin sources torsion vanishes, torsion does not propagate.

We have not included the cosmological constant in the field equations for simplicity. It should be noted, however, that there exist models where the cosmological constant might be induced by the torsion of spacetime. Likewise, torsion could contribute to the bare cosmological constant and yield today's observed effective cosmological term, see e.g.~\cite{Baekler:1987jb,Boehmer:2003iv,Yo:2006qs} and also~\cite{Cai:2008gk} for a spinorial dark energy model.

As for the matter we consider a dark spinor field. It obeys scalar field-like equations of motion since its mass dimension is one and the Lagrangian reads
\begin{align}
      \tilde{\lan} = \frac{1}{2} g^{ab} \tilde{\nabla}_{(a} \dual{\lambda}
      \tilde{\nabla}_{b)} \lambda - V(\dual{\lambda} \lambda),
      \label{eq:t0a}
\end{align}
where $V(\dual{\lambda}\lambda)$ denotes the dark spinor field potential. It is important to emphasize that the non-standard Wigner class spinors lead to more torsion structure than Dirac spinors~\cite{Boehmer:2006qq}. The resulting metric energy-momentum tensor and spin angular momentum tensor respectively become
\begin{align}
      \sigma_{ij} &= \tilde{\nabla}_{(i} \dual{\lambda} \tilde{\nabla}_{j)} \lambda - g_{ij} \tilde{\lan},\\
      \tau^{kj}{}_{i} &= \frac{1}{4} \tilde{\nabla}_i \dual{\lambda} \gamma^j \gamma^k \lambda - \frac{1}{4} \dual{\lambda} \gamma^j \gamma^k \tilde{\nabla}_i \lambda.
\end{align}

\section{Cosmological field equations}
\label{sec:cos}

Current observations~\cite{Riess:1998cb,Perlmutter:1998zf} suggest that the energy density of the universe is very close to the critical density, resulting in spatially flat hypersurfaces. The flat FLRW metric is
\begin{align}
      ds^2 = dt^2 - a(t)^2 \bigl( dx^2 + dy^2 + dz^2 \bigr),
      \label{eq:cos1a}
\end{align}
where $a(t)$ is the expansion parameter. It yields the following non-vanishing holonomic Christoffel symbols components
\begin{align}
      \Gamma_{tx}^{x} = \Gamma_{ty}^{y} = \Gamma_{tz}^{z} &= \frac{\dot{a}}{a} \,,
      \nonumber \\
      \Gamma_{xx}^{t} = \Gamma_{yy}^{t} = \Gamma_{zz}^{t} &= a\dot{a} \,,
      \label{eq:cos2}
\end{align}
where the dot denotes differentiation with respect to $t$. This then implies the following non-vanishing anholonomic Christoffel symbols $\Gamma_a$ to be
\begin{align}
      \Gamma_{n} &= -\frac{1}{2}\frac{\dot{a}}{a}
      (\gamma^{\hat{0}} \gamma^{n} - \gamma^{n} \gamma^{\hat{0}})
      = -2 \frac{\dot{a}}{a} f^{\hat{0}n}, \\
      n &= \hat{1},\hat{2},\hat{3}.
      \label{eq:cos3}
\end{align}
When the cosmological principle is applied to the torsion tensor~\cite{Tsamparlis:1979,Goenner:1984} the allowed components reduce to
\begin{align}
      T_{\hat{1}\hat{1}\hat{0}} &= T_{\hat{2}\hat{2}\hat{0}} =
      T_{\hat{3}\hat{3}\hat{0}} = h(t),\\
      T_{\hat{1}\hat{2}\hat{3}} &= T_{\hat{3}\hat{1}\hat{2}} =
      T_{\hat{2}\hat{3}\hat{1}} = f(t).
\end{align}
The cosmological Einstein tensor with torsion is now given by
\begin{align}
      G_{tt} &= 3\frac{\dot{a}}{a}\Bigl(\frac{\dot{a}}{a}+2h\Bigr) + 3h^2 - 3f^2,
      \label{gtt}\\
      G_{xx} &= a^2 \Bigl(-2\frac{\ddot{a}}{a}-\frac{\dot{a}}{a}
      \bigl(\frac{\dot{a}}{a}+4h\bigr)-2\dot{h}-h^2+f^2\Bigr),\\
      G_{xx} &= G_{yy} = G_{zz}.
      \label{gxx}
\end{align}

In addition to the geometry, also the matter has to be compatible with homogeneity and isotropy. This yields two classes of dark spinors, dark ghost spinors which satisfy $\dual{\lambda}\lambda = 0$ and standard dark spinors where $\dual{\lambda}\lambda \neq 0$. The name ghost spinors refers to the fact that such spinors lead to a vanishing metric energy-momentum tensor and hence do not effect the curvature of spacetime in general relativity, see also~\cite{Griffiths:1979,Griffiths:1981ym,Dimakis:1985jb,Boehmer:2007dh}. A cosmological ghost spinor field can be written in the form
\begin{align}
      \lambda_{\{-,+\}} = \varphi(t)\, \xi, \\
      \lambda_{\{+,-\}} = \varphi(t)\, \zeta,
      \label{eq:dy12}
\end{align}
where $\xi$ and $\zeta$ are two linearly independent constant spinors given by
\begin{align}
      \xi = \begin{pmatrix} 0\\ \pm i\\ 1\\ 0\end{pmatrix}, \qquad
      \zeta = i\begin{pmatrix} \mp i\\ 0\\ 0\\ -1\end{pmatrix},
      \label{eq:dy12app}
\end{align}
with their respective dual spinors
\begin{align}
      \gdual{\xi} &= i\begin{pmatrix} 0& i& \pm 1& 0\end{pmatrix},
      \nonumber \\
      \gdual{\zeta} &= \begin{pmatrix} -i& 0& 0& \mp 1\end{pmatrix}.
      \label{eq:dy13app}
\end{align}

The set of 24 algebraic equations~(\ref{eq:t9}) reduces to two independent equations relating spin and torsion if we assume homogeneity and isotropy. The torsion functions $f$ and $h$ can therefore be expressed~\footnote{These computations were performed using the software Mathematica} in terms of the matter
\begin{align}
      h = -\frac{\varphi^4/M_{\rm pl}^4}{4+\varphi^4/M_{\rm pl}^4}\,\frac{\dot{a}}{a},\\
      f = -\frac{2\varphi^2/M_{\rm pl}^2}{4+A_0^4/M_{\rm pl}^4}\,\frac{\dot{a}}{a},
      \label{eq:hf}
\end{align}
which can be combined to give
\begin{align}
      \frac{h}{f} = \frac{1}{2}\varphi^2/M_{\rm pl}^2.
      \label{eq:hf2}
\end{align}
Therefore, a dark ghost spinor field satisfying the cosmological principle indeed yields non-trivial contributions to the spatial axial torsion component and to the time component of the torsion vector. Hence, the spin angular momentum tensor induced by this matter source satisfies homogeneity and isotropy.

The total energy-momentum tensor $\Sigma_{ij}$ for the dark spinor matter is given by
\begin{align}
      \Sigma_{tt} &= V_0,
      \label{stt}\\
      \Sigma_{xx} &= -a^2 V_0 + a^2 \varphi^2
      \Bigl(3h - \frac{\dot{f}}{f} - 2\frac{\dot{\varphi}}{\varphi}\Bigr)f,
      \label{sxx} \\
      \Sigma_{xx} &= \Sigma_{yy}=\Sigma_{zz},
\end{align}
where $V_0 = V(0)$. This completes the formulation of the cosmological field equations. Next, we investigate the qualitative behavior of the equations of motion.

The geometrical part of the cosmological field equations~(\ref{gtt})--(\ref{gxx}) can, for example, be read off from~\cite{Goenner:1984} (cf their action $L_4$) which we verified. In Ref.~\cite{Minkowski:1986kv}, where $h=0$ was assumed, the geometry parameter $k$ was redefined to include the remaining torsion by $\bar{k}=k-f^2a^2/2$, see also~\cite{Boehmer:2005sw}.

\section{Cosmological dark spinor dynamics}
\label{sec:dyn}

The complete set of field equations can be reduced to a single first order differential equation in the following manner. First of all, all torsion functions in the field equations are written in terms of the spin tensor~(\ref{eq:hf}), thereby eliminating torsion $f$ and $h$ for the matter field $\varphi$. Next, we can use Eq.~(\ref{gtt}) and the derivative of that equation to find expressions for $\dot{a}/a$ and $\ddot{a}/a$ which are expressed entirely in terms of the matter field $\varphi$. We analyze these equations qualitatively and solve them numerically.

For the Hubble parameter $H=\dot{a}/a$ from Eq.~(\ref{gtt}) we find
\begin{align}
      H = \frac{\sqrt{V_0/M_{\rm pl}^2}}{2\sqrt{3}}
      \frac{4 + \varphi^4/M_{\rm pl}^4}{\sqrt{4 - \varphi^4/M_{\rm pl}^4}}.
      \label{eq:dy1}
\end{align}
Next, the terms with $\ddot{a}/a$, $\dot{a}/a$ and $f$ and $h$ are eliminated for $\varphi$ in the spatial component of the field equation which results in
\begin{align}
      \frac{\dot{\varphi}}{\varphi} = -\frac{\sqrt{V_0/M_{\rm pl}^2}}{4\sqrt{3}}
      \frac{8 + 3 \varphi^4/M_{\rm pl}^4}{12 -\varphi^4/M_{\rm pl}^4}
      \sqrt{4 - \varphi^4/M_{\rm pl}^4}.
      \label{eq:dy2}
\end{align}

Positivity of the square root requires $\varphi/M_{\rm pl} < \sqrt{2}$. This implies that the sign of the first derivative of the field cannot change its sign and hence the field value is a decreasing function of time and in fact quickly approaches zero. When this happens, the Hubble parameter asymptotes to a constant value and the universe expands according to $a \propto \exp(H t)$.

To see this behavior of the solutions qualitatively, let us expand Eqs.~(\ref{eq:dy1}) and~(\ref{eq:dy2}) about $\varphi=0$ which leads to
\begin{align}
      H &= \sqrt{\frac{V_0}{3M_{\rm pl}^2}} + O(\varphi/M_{\rm pl})^4,
      \label{eq:dy3}\\
      \frac{\dot{\varphi}}{\varphi} &= -\frac{1}{3}\sqrt{\frac{V_0}{3 M_{\rm pl}^2}} + 
      O(\varphi/M_{\rm pl})^4,
      \label{eq:dy4}
\end{align}
and therefore we find that a period of accelerated expansion is an attractor solution of this system of equations. Taking into account Eq.~(\ref{eq:hf}), we also find that the torsion of spacetime is quickly decreasing and approaching zero as the universe expands.

Such a behavior of the torsion is not unexpected, see e.g.~\cite{Bauerle:1983ai}. Spinors and inflation in the context of torsion theories have received much attention in the past~\cite{Gasperini:1986mv,Fennelly:1988dx,Chatterjee:1993rc,Obukhov:1993fd,Kao:1993nb,GarciadeAndrade:1999qt,Boehmer:2005sw}. It should be pointed out, however, that matter sources considered previously violate the cosmological principle.

\begin{figure}[!htb]
\centering
\includegraphics[width=0.48\textwidth]{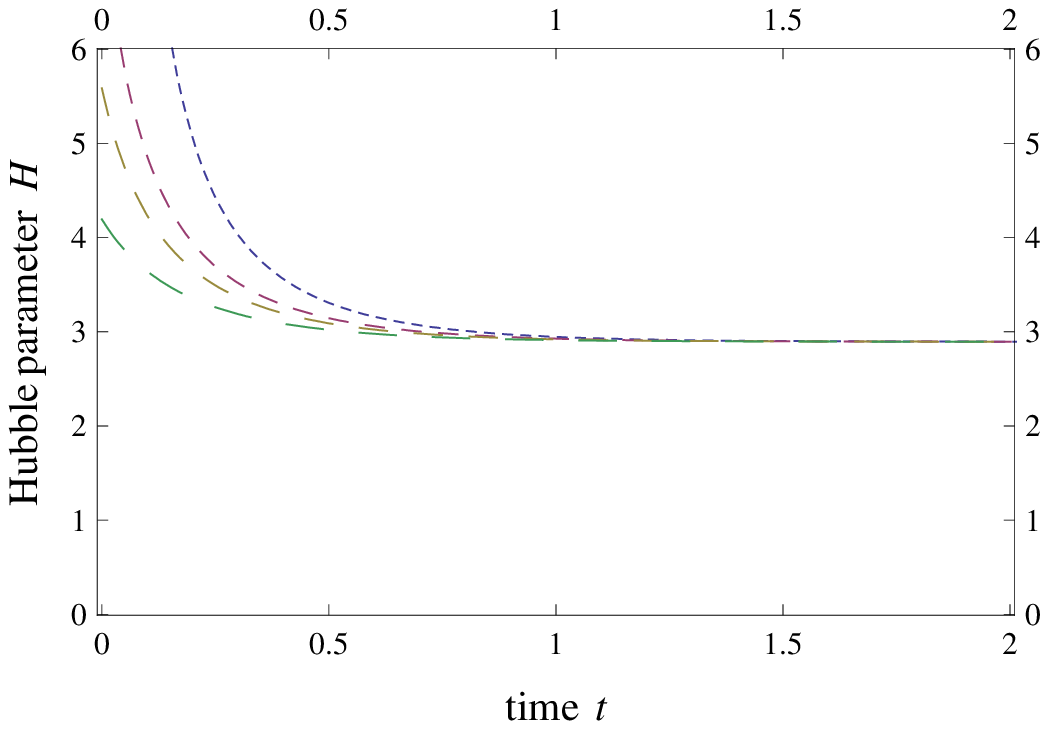}\hfill
\includegraphics[width=0.48\textwidth]{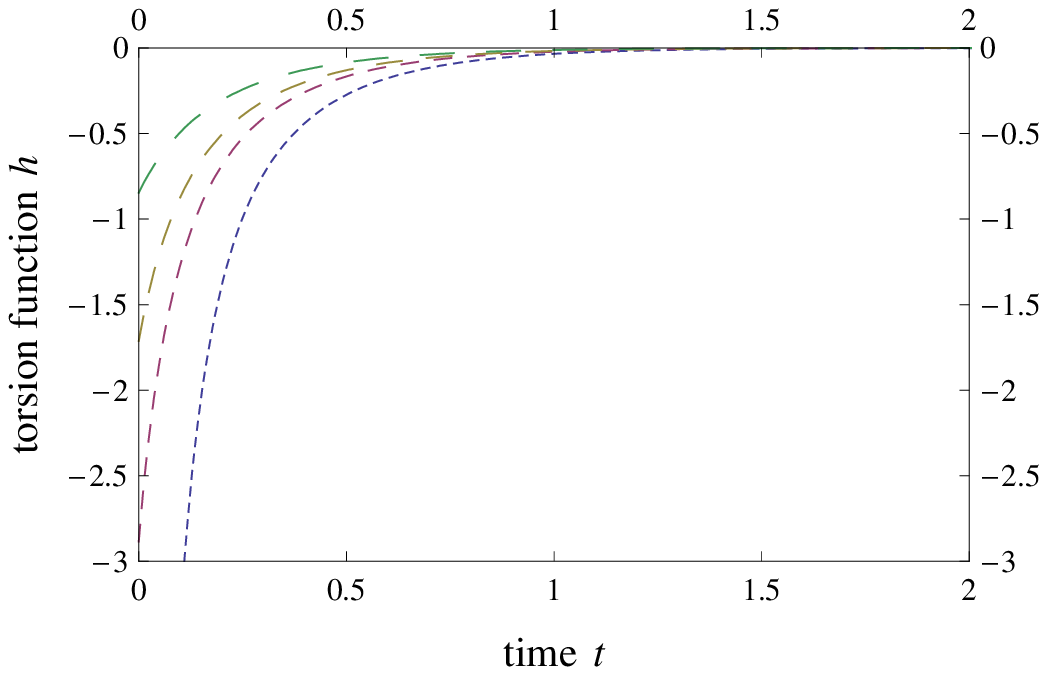}
\caption{Left: Hubble parameter and right: torsion function $h$ for $1/M_{\rm pl}^2 = 8\pi$ and $V_0 = 1$. Initial conditions of the matter field are $\varphi_i=\varphi(t=0)=\{0.282,0.25,0.23,0.20\}$, $\{$blue (short dashed), red (dashed), (medium dashed) yellow and green (long dashed) $\}$}.
\label{fig1}
\end{figure}

We numerically solve the first order differential equation~(\ref{eq:dy2}) and use this solution to find the evolution of the Hubble parameter, we plot the Hubble parameter in Fig.~\ref{fig1}a, which approaches to a constant for different initial conditions of the field. In Fig.~\ref{fig1}b the torsion function $h$ is plotted for the same numerical solutions.

\begin{figure}[!htb]
\centering
\includegraphics[width=0.48\textwidth]{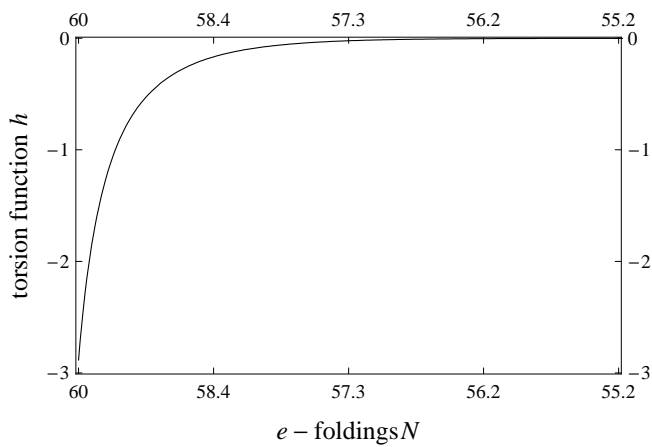}
\caption{Torsion function $h$ for $1/M_{\rm pl}^2 = 8\pi$ and $V_0 = 1$. Initial condition is $\varphi_i=\{0.25\}$.}
\label{fig3}
\end{figure}

In order to give a qualitative statement about the decay rate of the torsion, in Fig.~\ref{fig3} we plot the torsion function $h$ as a function of the number of $e$-foldings. We assume the total number of $e$-foldings to be sixty.
Therefore, the torsion contribution of the spacetime becomes negligible after approximately four $e$-foldings.

\section{Conclusions}
\label{sec:con}

We identified a matter source whose spin-angular momentum tensor is compatible with the cosmological principle. We then solved the resulting field equations of Einstein-Cartan theory. This matter source consists of the eigenspinors of the charge conjugation operator, which we refer to as a dark spinor field. It couples to all irreducible parts of torsion and therefore leads to an interesting coupling of matter and geometry. This matter source is also naturally dark in that it can only interact via the Higgs mechanism or gravity.

Our solutions of the field equations show that torsion does vanish quickly (approximately after a few e-foldings) and that the Hubble parameter has a constant value as an attractor. Both features of the model fit very well into the standard model of inflationary cosmology in that a period of accelerated expansion is an attractor solution. It is worth noting that in Einstein-Cartan theory the spins of elementary particles are thought to be the primary sources of torsion, and it is therefore expected that on large sales and over time torsion should average out or decay, respectively.

We speculate that some non-zero cosmological torsion has already been observed in the large scale anisotropies of the cosmic microwave background radiation (CMB) where torsion leaves its imprint only on the largest scales.

\acknowledgments
We thank Friedrich Hehl and Dmitri Vassiliev for discussions.


\begin{thebibliography}{45}
\expandafter\ifx\csname natexlab\endcsname\relax\def\natexlab#1{#1}\fi
\expandafter\ifx\csname bibnamefont\endcsname\relax
  \def\bibnamefont#1{#1}\fi
\expandafter\ifx\csname bibfnamefont\endcsname\relax
  \def\bibfnamefont#1{#1}\fi
\expandafter\ifx\csname citenamefont\endcsname\relax
  \def\citenamefont#1{#1}\fi
\expandafter\ifx\csname url\endcsname\relax
  \def\url#1{\texttt{#1}}\fi
\expandafter\ifx\csname urlprefix\endcsname\relax\def\urlprefix{URL }\fi
\providecommand{\bibinfo}[2]{#2}
\providecommand{\eprint}[2][]{\url{#2}}

\bibitem[{\citenamefont{Hehl}(1973)}]{Hehl:1973}
\bibinfo{author}{\bibfnamefont{F.~W.} \bibnamefont{Hehl}},
  \bibinfo{journal}{Gen. Rel. Grav.} \textbf{\bibinfo{volume}{4}},
  \bibinfo{pages}{333} (\bibinfo{year}{1973}).

\bibitem[{\citenamefont{Hehl}(1974)}]{Hehl:1974}
\bibinfo{author}{\bibfnamefont{F.~W.} \bibnamefont{Hehl}},
  \bibinfo{journal}{Gen. Rel. Grav.} \textbf{\bibinfo{volume}{5}},
  \bibinfo{pages}{491} (\bibinfo{year}{1974}).

\bibitem[{\citenamefont{Hehl et~al.}(1976)\citenamefont{Hehl, von~der Heyde,
  Kerlick, and Nester}}]{Hehl:1976kj}
\bibinfo{author}{\bibfnamefont{F.~W.} \bibnamefont{Hehl}},
  \bibinfo{author}{\bibfnamefont{P.}~\bibnamefont{von~der Heyde}},
  \bibinfo{author}{\bibfnamefont{G.~D.} \bibnamefont{Kerlick}},
  \bibnamefont{and} \bibinfo{author}{\bibfnamefont{J.~M.}
  \bibnamefont{Nester}}, \bibinfo{journal}{Rev. Mod. Phys.}
  \textbf{\bibinfo{volume}{48}}, \bibinfo{pages}{393} (\bibinfo{year}{1976}).

\bibitem[{\citenamefont{Hehl et~al.}(1995)\citenamefont{Hehl, McCrea, Mielke,
  and Neeman}}]{Hehl:1995ue}
\bibinfo{author}{\bibfnamefont{F.~W.} \bibnamefont{Hehl}},
  \bibinfo{author}{\bibfnamefont{J.~D.} \bibnamefont{McCrea}},
  \bibinfo{author}{\bibfnamefont{E.~W.} \bibnamefont{Mielke}},
  \bibnamefont{and} \bibinfo{author}{\bibfnamefont{Y.}~\bibnamefont{Neeman}},
  \bibinfo{journal}{Phys. Rept.} \textbf{\bibinfo{volume}{258}},
  \bibinfo{pages}{1} (\bibinfo{year}{1995}), \eprint{gr-qc/9402012}.

\bibitem[{\citenamefont{Hammond}(2002)}]{Hammond:2002rm}
\bibinfo{author}{\bibfnamefont{R.~T.} \bibnamefont{Hammond}},
  \bibinfo{journal}{Rept. Prog. Phys.} \textbf{\bibinfo{volume}{65}},
  \bibinfo{pages}{599} (\bibinfo{year}{2002}).

\bibitem[{\citenamefont{Trautman}(2006)}]{Trautman:2006fp}
\bibinfo{author}{\bibfnamefont{A.}~\bibnamefont{Trautman}},
  \bibinfo{journal}{Encyclopedia of Mathematical Physics}
  \textbf{\bibinfo{volume}{2}}, \bibinfo{pages}{189} (\bibinfo{year}{2006}),
  \eprint{gr-qc/0606062}.

\bibitem[{\citenamefont{Tsamparlis}(1979)}]{Tsamparlis:1979}
\bibinfo{author}{\bibfnamefont{M.}~\bibnamefont{Tsamparlis}},
  \bibinfo{journal}{Phys. Lett.} \textbf{\bibinfo{volume}{75A}},
  \bibinfo{pages}{27} (\bibinfo{year}{1979}).

\bibitem[{\citenamefont{Kopczy{\'n}ski}(1972)}]{Kopczynski:1972}
\bibinfo{author}{\bibfnamefont{W.}~\bibnamefont{Kopczy{\'n}ski}},
  \bibinfo{journal}{Phys. Lett.} \textbf{\bibinfo{volume}{A39}},
  \bibinfo{pages}{219} (\bibinfo{year}{1972}).

\bibitem[{\citenamefont{Kopczy{\'n}ski}(1973)}]{Kopczynski:1973}
\bibinfo{author}{\bibfnamefont{W.}~\bibnamefont{Kopczy{\'n}ski}},
  \bibinfo{journal}{Phys. Lett.} \textbf{\bibinfo{volume}{A43}},
  \bibinfo{pages}{63} (\bibinfo{year}{1973}).

\bibitem[{\citenamefont{Weyssenhoff and Raabe}(1947)}]{Weyssenhoff:1947}
\bibinfo{author}{\bibfnamefont{J.}~\bibnamefont{Weyssenhoff}} \bibnamefont{and}
  \bibinfo{author}{\bibfnamefont{A.}~\bibnamefont{Raabe}},
  \bibinfo{journal}{Acta Phys. Pol.} \textbf{\bibinfo{volume}{IX}},
  \bibinfo{pages}{7} (\bibinfo{year}{1947}).

\bibitem[{\citenamefont{Obukhov and Korotkii}(1987)}]{Obukhov:1987yu}
\bibinfo{author}{\bibfnamefont{Y.~N.} \bibnamefont{Obukhov}} \bibnamefont{and}
  \bibinfo{author}{\bibfnamefont{V.~A.} \bibnamefont{Korotkii}},
  \bibinfo{journal}{Class. Quant. Grav.} \textbf{\bibinfo{volume}{4}},
  \bibinfo{pages}{1633} (\bibinfo{year}{1987}).

\bibitem[{\citenamefont{B{\"o}hmer and Bronowski}(2006)}]{Boehmer:2006gd}
\bibinfo{author}{\bibfnamefont{C.~G.} \bibnamefont{B{\"o}hmer}}
  \bibnamefont{and} \bibinfo{author}{\bibfnamefont{P.}~\bibnamefont{Bronowski}}
  (\bibinfo{year}{2006}), \eprint{gr-qc/0601089}.

\bibitem[{\citenamefont{Brechet et~al.}(2008)\citenamefont{Brechet, Hobson, and
  Lasenby}}]{Brechet:2008tz}
\bibinfo{author}{\bibfnamefont{S.~D.} \bibnamefont{Brechet}},
  \bibinfo{author}{\bibfnamefont{M.~P.} \bibnamefont{Hobson}},
  \bibnamefont{and} \bibinfo{author}{\bibfnamefont{A.~N.}
  \bibnamefont{Lasenby}} (\bibinfo{year}{2008}), \eprint{0807.2523}.

\bibitem[{\citenamefont{Goenner and M{\"u}ller-Hoissen}(1984)}]{Goenner:1984}
\bibinfo{author}{\bibfnamefont{H.~F.~M.} \bibnamefont{Goenner}}
  \bibnamefont{and}
  \bibinfo{author}{\bibfnamefont{F.}~\bibnamefont{M{\"u}ller-Hoissen}},
  \bibinfo{journal}{Class. Quant. Grav.} \textbf{\bibinfo{volume}{1}},
  \bibinfo{pages}{651} (\bibinfo{year}{1984}).

\bibitem[{\citenamefont{Lasenby et~al.}(2005)\citenamefont{Lasenby, Doran, and
  Heineke}}]{Lasenby:2005nw}
\bibinfo{author}{\bibfnamefont{A.}~\bibnamefont{Lasenby}},
  \bibinfo{author}{\bibfnamefont{C.~J.~L.} \bibnamefont{Doran}},
  \bibnamefont{and} \bibinfo{author}{\bibfnamefont{R.}~\bibnamefont{Heineke}}
  (\bibinfo{year}{2005}), \eprint{gr-qc/0509014}.

\bibitem[{\citenamefont{Puetzfeld and Chen}(2004)}]{Puetzfeld:2004df}
\bibinfo{author}{\bibfnamefont{D.}~\bibnamefont{Puetzfeld}} \bibnamefont{and}
  \bibinfo{author}{\bibfnamefont{X.-L.} \bibnamefont{Chen}},
  \bibinfo{journal}{Class. Quant. Grav.} \textbf{\bibinfo{volume}{21}},
  \bibinfo{pages}{2703} (\bibinfo{year}{2004}), \eprint{gr-qc/0402026}.

\bibitem[{\citenamefont{Puetzfeld}(2005)}]{Puetzfeld:2004yg}
\bibinfo{author}{\bibfnamefont{D.}~\bibnamefont{Puetzfeld}},
  \bibinfo{journal}{New Astron. Rev.} \textbf{\bibinfo{volume}{49}},
  \bibinfo{pages}{59} (\bibinfo{year}{2005}), \eprint{gr-qc/0404119}.

\bibitem[{\citenamefont{Puetzfeld et~al.}(2005)\citenamefont{Puetzfeld, Pohl,
  and Zhu}}]{Puetzfeld:2004sw}
\bibinfo{author}{\bibfnamefont{D.}~\bibnamefont{Puetzfeld}},
  \bibinfo{author}{\bibfnamefont{M.}~\bibnamefont{Pohl}}, \bibnamefont{and}
  \bibinfo{author}{\bibfnamefont{Z.-H.} \bibnamefont{Zhu}},
  \bibinfo{journal}{Astrophys. J.} \textbf{\bibinfo{volume}{619}},
  \bibinfo{pages}{657} (\bibinfo{year}{2005}), \eprint{astro-ph/0407204}.

\bibitem[{\citenamefont{Puetzfeld and Obukhov}(2007)}]{Puetzfeld:2007ye}
\bibinfo{author}{\bibfnamefont{D.}~\bibnamefont{Puetzfeld}} \bibnamefont{and}
  \bibinfo{author}{\bibfnamefont{Y.~N.} \bibnamefont{Obukhov}}
  (\bibinfo{year}{2007}), \eprint{arXiv:0708.1926 [gr-qc]}.

\bibitem[{\citenamefont{Ahluwalia-Khalilova and
  Grumiller}(2005{\natexlab{a}})}]{jcap}
\bibinfo{author}{\bibfnamefont{D.~V.} \bibnamefont{Ahluwalia-Khalilova}}
  \bibnamefont{and}
  \bibinfo{author}{\bibfnamefont{D.}~\bibnamefont{Grumiller}},
  \bibinfo{journal}{JCAP} \textbf{\bibinfo{volume}{0507}}, \bibinfo{pages}{012}
  (\bibinfo{year}{2005}{\natexlab{a}}), \eprint{hep-th/0412080}.

\bibitem[{\citenamefont{Ahluwalia-Khalilova and
  Grumiller}(2005{\natexlab{b}})}]{prd}
\bibinfo{author}{\bibfnamefont{D.~V.} \bibnamefont{Ahluwalia-Khalilova}}
  \bibnamefont{and}
  \bibinfo{author}{\bibfnamefont{D.}~\bibnamefont{Grumiller}},
  \bibinfo{journal}{Phys. Rev.} \textbf{\bibinfo{volume}{D72}},
  \bibinfo{pages}{067701} (\bibinfo{year}{2005}{\natexlab{b}}),
  \eprint{hep-th/0410192}.

\bibitem[{\citenamefont{da~Rocha and Rodrigues}(2006)}]{daRocha:2005ti}
\bibinfo{author}{\bibfnamefont{R.}~\bibnamefont{da~Rocha}} \bibnamefont{and}
  \bibinfo{author}{\bibfnamefont{J.}~\bibnamefont{Rodrigues},
  \bibfnamefont{W.~A.}}, \bibinfo{journal}{Mod. Phys. Lett.}
  \textbf{\bibinfo{volume}{A21}}, \bibinfo{pages}{65} (\bibinfo{year}{2006}),
  \eprint{math-ph/0506075}.

\bibitem[{\citenamefont{B{\"o}hmer}(2007{\natexlab{a}})}]{Boehmer:2006qq}
\bibinfo{author}{\bibfnamefont{C.~G.} \bibnamefont{B{\"o}hmer}},
  \bibinfo{journal}{Annalen Phys. (Leipzig)} \textbf{\bibinfo{volume}{16}},
  \bibinfo{pages}{38} (\bibinfo{year}{2007}{\natexlab{a}}),
  \eprint{gr-qc/0607088}.

\bibitem[{\citenamefont{B{\"o}hmer and Mota}(2008)}]{Boehmer:2007ut}
\bibinfo{author}{\bibfnamefont{C.~G.} \bibnamefont{B{\"o}hmer}}
  \bibnamefont{and} \bibinfo{author}{\bibfnamefont{D.~F.} \bibnamefont{Mota}},
  \bibinfo{journal}{Phys. Lett.} \textbf{\bibinfo{volume}{B663}},
  \bibinfo{pages}{168} (\bibinfo{year}{2008}), \eprint{arXiv:0710.2003
  [astro-ph]}.

\bibitem[{\citenamefont{B{\"o}hmer}(2008)}]{Boehmer:2008rz}
\bibinfo{author}{\bibfnamefont{C.~G.} \bibnamefont{B{\"o}hmer}},
  \bibinfo{journal}{Phys. Rev.} \textbf{\bibinfo{volume}{D77}},
  \bibinfo{pages}{123535} (\bibinfo{year}{2008}), \eprint{arXiv:0804.0616
  [astro-ph]}.

\bibitem[{\citenamefont{Gredat and Shankaranarayanan}(2008)}]{Gredat:2008qf}
\bibinfo{author}{\bibfnamefont{D.}~\bibnamefont{Gredat}} \bibnamefont{and}
  \bibinfo{author}{\bibfnamefont{S.}~\bibnamefont{Shankaranarayanan}}
  (\bibinfo{year}{2008}), \eprint{arXiv:0807.3336 [astro-ph]}.

\bibitem[{\citenamefont{Baekler et~al.}(1987)\citenamefont{Baekler, Mielke,
  Hecht, and Hehl}}]{Baekler:1987jb}
\bibinfo{author}{\bibfnamefont{P.}~\bibnamefont{Baekler}},
  \bibinfo{author}{\bibfnamefont{E.~W.} \bibnamefont{Mielke}},
  \bibinfo{author}{\bibfnamefont{R.}~\bibnamefont{Hecht}}, \bibnamefont{and}
  \bibinfo{author}{\bibfnamefont{F.~W.} \bibnamefont{Hehl}},
  \bibinfo{journal}{Nucl. Phys.} \textbf{\bibinfo{volume}{B288}},
  \bibinfo{pages}{800} (\bibinfo{year}{1987}).

\bibitem[{\citenamefont{B{\"o}hmer}(2004)}]{Boehmer:2003iv}
\bibinfo{author}{\bibfnamefont{C.~G.} \bibnamefont{B{\"o}hmer}},
  \bibinfo{journal}{Class. Quant. Grav.} \textbf{\bibinfo{volume}{21}},
  \bibinfo{pages}{1119} (\bibinfo{year}{2004}), \eprint{gr-qc/0310058}.

\bibitem[{\citenamefont{Yo and Nester}(2006)}]{Yo:2006qs}
\bibinfo{author}{\bibfnamefont{H.-J.} \bibnamefont{Yo}} \bibnamefont{and}
  \bibinfo{author}{\bibfnamefont{J.~M.} \bibnamefont{Nester}}
  (\bibinfo{year}{2006}), \eprint{astro-ph/0612738}.

\bibitem[{\citenamefont{Cai and Wang}(2008)}]{Cai:2008gk}
\bibinfo{author}{\bibfnamefont{Y.-F.} \bibnamefont{Cai}} \bibnamefont{and}
  \bibinfo{author}{\bibfnamefont{J.}~\bibnamefont{Wang}},
  \bibinfo{journal}{Class. Quant. Grav.} \textbf{\bibinfo{volume}{25}},
  \bibinfo{pages}{165014} (\bibinfo{year}{2008}), 
  \eprint{arXiv:0806.3890 [hep-th]}.

\bibitem[{\citenamefont{Riess et~al.}(1998)}]{Riess:1998cb}
\bibinfo{author}{\bibfnamefont{A.~G.} \bibnamefont{Riess}} \bibnamefont{et~al.}
  (\bibinfo{collaboration}{Supernova Search Team}), \bibinfo{journal}{Astron.
  J.} \textbf{\bibinfo{volume}{116}}, \bibinfo{pages}{1009}
  (\bibinfo{year}{1998}), \eprint{astro-ph/9805201}.

\bibitem[{\citenamefont{Perlmutter et~al.}(1998)}]{Perlmutter:1998zf}
\bibinfo{author}{\bibfnamefont{S.}~\bibnamefont{Perlmutter}}
  \bibnamefont{et~al.} (\bibinfo{collaboration}{Supernova Cosmology Project}),
  \bibinfo{journal}{Nature} \textbf{\bibinfo{volume}{391}}, \bibinfo{pages}{51}
  (\bibinfo{year}{1998}), \eprint{astro-ph/9712212}.

\bibitem[{\citenamefont{Griffiths}(1979)}]{Griffiths:1979}
\bibinfo{author}{\bibfnamefont{J.~B.} \bibnamefont{Griffiths}},
  \bibinfo{journal}{J. Phy.} \textbf{\bibinfo{volume}{A12}},
  \bibinfo{pages}{2429} (\bibinfo{year}{1979}).

\bibitem[{\citenamefont{Griffiths}(1981)}]{Griffiths:1981ym}
\bibinfo{author}{\bibfnamefont{J.~B.} \bibnamefont{Griffiths}},
  \bibinfo{journal}{Gen. Rel. Grav.} \textbf{\bibinfo{volume}{13}},
  \bibinfo{pages}{227} (\bibinfo{year}{1981}).

\bibitem[{\citenamefont{Dimakis and M{\"u}ller-Hoissen}(1985)}]{Dimakis:1985jb}
\bibinfo{author}{\bibfnamefont{A.}~\bibnamefont{Dimakis}} \bibnamefont{and}
  \bibinfo{author}{\bibfnamefont{F.}~\bibnamefont{M{\"u}ller-Hoissen}},
  \bibinfo{journal}{J. Math. Phys.} \textbf{\bibinfo{volume}{26}},
  \bibinfo{pages}{1040} (\bibinfo{year}{1985}).

\bibitem[{\citenamefont{B{\"o}hmer}(2007{\natexlab{b}})}]{Boehmer:2007dh}
\bibinfo{author}{\bibfnamefont{C.~G.} \bibnamefont{B{\"o}hmer}},
  \bibinfo{journal}{Annalen Phys. (Leipzig)} \textbf{\bibinfo{volume}{16}},
  \bibinfo{pages}{325} (\bibinfo{year}{2007}{\natexlab{b}}),
  \eprint{gr-qc/0701087}.

\bibitem[{\citenamefont{Minkowski}(1986)}]{Minkowski:1986kv}
\bibinfo{author}{\bibfnamefont{P.}~\bibnamefont{Minkowski}},
  \bibinfo{journal}{Phys. Lett.} \textbf{\bibinfo{volume}{B173}},
  \bibinfo{pages}{247} (\bibinfo{year}{1986}).

\bibitem[{\citenamefont{B{\"o}hmer}(2005)}]{Boehmer:2005sw}
\bibinfo{author}{\bibfnamefont{C.~G.} \bibnamefont{B{\"o}hmer}},
  \bibinfo{journal}{Acta Phys. Polon.} \textbf{\bibinfo{volume}{B36}},
  \bibinfo{pages}{2841} (\bibinfo{year}{2005}), \eprint{gr-qc/0506033}.

\bibitem[{\citenamefont{B{\"a}uerle and Haneveld}(1983)}]{Bauerle:1983ai}
\bibinfo{author}{\bibfnamefont{G.~G.~A.} \bibnamefont{B{\"a}uerle}}
  \bibnamefont{and} \bibinfo{author}{\bibfnamefont{C.~J.}
  \bibnamefont{Haneveld}}, \bibinfo{journal}{Physica}
  \textbf{\bibinfo{volume}{121A}}, \bibinfo{pages}{541} (\bibinfo{year}{1983}).

\bibitem[{\citenamefont{Gasperini}(1986)}]{Gasperini:1986mv}
\bibinfo{author}{\bibfnamefont{M.}~\bibnamefont{Gasperini}},
  \bibinfo{journal}{Phys. Rev. Lett.} \textbf{\bibinfo{volume}{56}},
  \bibinfo{pages}{2873} (\bibinfo{year}{1986}).

\bibitem[{\citenamefont{Fennelly et~al.}(1988)\citenamefont{Fennelly, Bradas,
  and Smalley}}]{Fennelly:1988dx}
\bibinfo{author}{\bibfnamefont{A.~J.} \bibnamefont{Fennelly}},
  \bibinfo{author}{\bibfnamefont{J.~C.} \bibnamefont{Bradas}},
  \bibnamefont{and} \bibinfo{author}{\bibfnamefont{L.~L.}
  \bibnamefont{Smalley}}, \bibinfo{journal}{Phys. Lett.}
  \textbf{\bibinfo{volume}{A129}}, \bibinfo{pages}{195} (\bibinfo{year}{1988}).

\bibitem[{\citenamefont{Chatterjee and Bhattacharya}(1993)}]{Chatterjee:1993rc}
\bibinfo{author}{\bibfnamefont{P.}~\bibnamefont{Chatterjee}} \bibnamefont{and}
  \bibinfo{author}{\bibfnamefont{B.}~\bibnamefont{Bhattacharya}},
  \bibinfo{journal}{Mod. Phys. Lett.} \textbf{\bibinfo{volume}{A8}},
  \bibinfo{pages}{2249} (\bibinfo{year}{1993}).

\bibitem[{\citenamefont{Obukhov}(1993)}]{Obukhov:1993fd}
\bibinfo{author}{\bibfnamefont{Y.~N.} \bibnamefont{Obukhov}},
  \bibinfo{journal}{Phys. Lett.} \textbf{\bibinfo{volume}{A182}},
  \bibinfo{pages}{214} (\bibinfo{year}{1993}), \eprint{gr-qc/0008015}.

\bibitem[{\citenamefont{Kao}(1993)}]{Kao:1993nb}
\bibinfo{author}{\bibfnamefont{W.~F.} \bibnamefont{Kao}},
  \bibinfo{journal}{Phys. Rev.} \textbf{\bibinfo{volume}{D47}},
  \bibinfo{pages}{3639} (\bibinfo{year}{1993}).

\bibitem[{\citenamefont{Garcia~de Andrade}(1999)}]{GarciadeAndrade:1999qt}
\bibinfo{author}{\bibfnamefont{L.~C.} \bibnamefont{Garcia~de Andrade}},
  \bibinfo{journal}{Int. J. Mod. Phys.} \textbf{\bibinfo{volume}{D8}},
  \bibinfo{pages}{725} (\bibinfo{year}{1999}).

\end{thebibliography}
\end{document}